# Trillion-atom molecular dynamics simulations with *ab initio* accuracy

Pengfei Suo[1,*], Wudi Cao[1,*], Xingxing Wu[1,*], Wenjie Zhang[2], Zheyong Fan[2,6], Shuanghan Xian[3], Rui Wang[3], Cheng Qian[2], Chao Liang[3], Qinghong Yuan[2,7], Xiaoshuang Chen[2,8], Pengfei Guan[4], Jingde Bu[3], Hongzhen Tian[1,§], Yanjing Su[2,9,§], Feng Ding[2,§] and Lin-Wang Wang[5,§]

1 Technology Department, Beijing Lonxun Quantum Co., Ltd., Beijing, P. R. China
2 Suzhou Laboratory, Suzhou, Jiangsu, P. R. China
3 National High Performance Computer Engineering Technology Research Center, Beijing, P. R. China
4 Ningbo Institute of Materials Technology and Engineering, Chinese Academy of Sciences, Ningbo, P. R. China
5 State Key Laboratory of Optoelectronic Materials and Devices, Institute of Semiconductors, Chinese Academy of Science, Beijing, P. R. China
6 College of Physical Science and Technology, Bohai University, Jinzhou, P. R. China
7 School of Physics and Electronic Science, East China Normal University, Shanghai, P. R. China
8 Shanghai Institute of Technical Physics, Chinese Academy of Sciences, Shanghai, P. R. China
9 Beijing Advanced Innovation Center for Materials Genome Engineering, University of Science and Technology Beijing, Beijing, P. R. China

*Abstract*—Material properties are fundamentally dictated by multiscale phenomena, which often reach mesoscale in size. The μm mesoscale is also the size which can be observed directly under an optical microscope, bridging the atomistic microscopic description with the continuous model macroscopic world. In this work, we report an unprecedented molecular dynamics (MD) simulation comprising 1.62 trillion atoms. Utilizing the Neuroevolution Potential (NEP) framework, we attained *ab initio* accuracy on China's New-generation Intelligent Supercomputer. Our implementation achieves a time-to-solution (s/step/atom) 100 times faster than previous state-of-the-art machine learning force field simulations, and 1,000 times faster than the Gordon Bell Prize-winning application from six years ago. Furthermore, we demonstrate an 86.9% weak scaling efficiency from a single GPGPU to 45,000 GPGPUs. These results redefine atomistic simulation boundaries, enabling direct mesoscopic modeling with quantum-level precision.

*Keywords*

neuroevolution potential, molecular dynamics, machine learning, GPU, MatPL, trillion atoms simulation, mesoscale

## I. Justification for ACM Gordon Bell Prize

World's first trillion-atom molecular dynamics simulation at *ab initio* accuracy, reached 1.62 trillion atoms at the 2.26×2.26×3.77 μm³ mesoscale, with time-to-solution of 5.0×10⁻¹³ s/step/atom, 100× scale-up over prior state-of-the-art and 1,000× over the 2020 Gordon Bell Prize-winning DeePMD work. For the first time, "trillion atoms" becomes reality in first-principles-accurate simulation.

*Pengfei Suo, Wudi Cao, Xingxing Wu contributed equally to this work
§Correspondence: tianhongzhen@pwmat.com, yjsu@ustb.edu.cn, dingf@szlab.ac.cn, lwwang@semi.ac.cn

## II. Performance Attributes

| Performance attribute | Our submission |
| --- | --- |
| Category of achievement | Time-to-solution, scalability |
| Type of method used | NEP, molecular dynamics |
| Results reported on basis of | Whole application, except I/O |
| Precision reported | mixed precision |
| System scale | Results measured on full-scale system |
| Measurements | Timers, FLOP count |

## III. Overview of the Problem

### A. Mesoscale atomistic molecular dynamics simulations

Pushing the boundaries of system size remains a primary pursuit in computational materials science, largely driven by the multiscale nature of materials properties. For instance, the macroscopic mechanical behavior of a metal is dictated by microscopic dislocation dynamics and pinning from atomic precipitations [1]. Similarly, the degradation of lithium battery cathodes manifests as the fracturing of μm-sized particles, yet this macroscopic failure is strictly initiated by atomic-scale defects and localized stress [2]. Capturing these multiscale phenomena is critical. In most scenarios, these complex behaviors converge and stabilize at the mesoscale - a spatial regime spanning several micrometers that serves as the critical bridge between atomistic interactions and macroscopic properties [3]. Once a system reaches the mesoscale, its physical behaviors approach bulk limits, allowing for accurate descriptions via phenomenological parameters and continuum models. Therefore, achieving direct, atomistic-level simulations at the mesoscale represents a highly coveted milestone in the field.

While it is computationally inexpensive to simulate large systems using simple empirical models like the Lennard-Jones (LJ) potential, describing complex physical phenomena with confidence strictly requires accurate machine learning force fields (MLFF) [4–7] trained on *ab initio* data. To the best of our knowledge, the largest prior simulation using an MLFF contains 29 billion atoms with about 2 seconds per MD step (Table. 1), while the Gordon Bell prize winning work 6 years ago only used 100 million atoms with about 0.1 second per MD step. Crucially, these benchmarks fall short of the true mesoscale, especially when a rapid time-to-solution is required. In the current work, we report an MD simulation with over trillion atoms simulations. It reaches a time-to-solution of $5.0\times10^{-13}$ s/step/atom (the smaller the better), representing $100\times$ and $1000\times$ increases compared to the previous 29B atoms and 2020 Gordon Bell prize works respectively. The 1.6 trillion Cu atom system represents $2.26\times2.26\times3.77$ $\mu m^3$ in size, it is thus a true mesoscopic system. Our work is achieved by three techniques. First, we have improved the MLFF. Unlike the previous MLFF which relies on deep neural network, we have used a light weight neuroevolution potential (NEP) approach [8], which is based on well-designed atom-environment features and a neural network with a single hidden layer. Nevertheless, trained for a given physics problem, such relatively shallow neural network is flexible enough to have similar accuracy as the deep neural network model while 100 times faster in inference speed. This can be done using our newly developed MatPL package [9], which has a special training and sampling procedure. The smaller NEP model also requires much smaller memory thus allowing much larger system simulation. Second, we have implemented our MD code on the China's New-generation Intelligent Supercomputer (hereinafter referred to as CNIS). We have taken the full advantage of its tensor-core feature, which can speed up the neural network calculations. In particular, batched calculations of multiple atoms are implemented on each GPGPU core, transforming the neural network computation into highly efficient matrix-matrix multiplications. All these optimizations yield a $9\times$ speedup over our initial implementations. Third, to ensure broad usability, we integrated our NEP model into the widely used LAMMPS [10] framework. This seamlessly allows users to leverage existing LAMMPS simulation protocols, while we capitalize on its robust spatial decomposition strategies, including ghost atom buffering and optimized communication patterns.

*B. Neuroevolution potential*

MLFFs offer a pathway to achieve computational accuracy at the level of the underlying training data. Starting from the foundational high-dimensional neural network potential by Behler and Parrinello [11], which features a computational cost scaling linearly with system size, many different MLFF approaches have been proposed. These include, for example, the Gaussian approximation potential (GAP) [12], the spectral neighbor analysis potential (SNAP) [13], the deep potential (DP) [14], the neuroevolution potential (NEP) [8], and the Allegro potential [15]. Most of these have demonstrated high computational performance [16–18]. Here, we employ the NEP approach, which is scalable to many-species systems [19] and has enabled high-fidelity simulations of a wide range of materials and properties [20, 21].

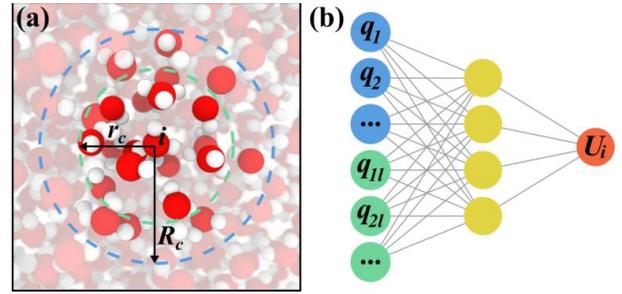

Fig. 1. Schematic illustration of the NEP architecture. (a) The dashed circles indicate the different cutoff radii for the radial and angular descriptors, denoted $R_c$ and $r_c$, respectively. (b) The input layer comprising radial ($q_n$) and angular ($q_{nl}$) descriptors is mapped to the atomic energy $U_i$ through a hidden layer.

As shown in Fig. 1, NEP adopts a fully connected neural network architecture. For each atom $i$ in the simulated system, its potential energy $U_i$ is expressed as the output of the same neural network function, but with possibly different weight and bias parameters depending on the atomic species. The input layer consists of an array of $N_{des}$ atom-environment descriptors $\mathbf{q}_i$ for atom $i$. The total potential energy of the system $U$ is the sum of the atomic energies. In the simplest case, where there is only one hidden layer with $N_{neu}$ neurons, which is the case for the NEP models used in the present work, the atomic energy can be explicitly expressed as

$$U_i = \sum_{\mu=1}^{N_{neu}} w_\mu^{(1)} \tanh\left(\sum_{\nu=1}^{N_{des}} w_{\mu\nu}^{(0)} q_\nu^i - b_\mu^{(0)}\right) - b^{(1)},$$

where $\mathbf{w}$ and $\mathbf{b}$ are the trainable weight and bias parameters, and $\tanh(x)$ is the hyperbolic tangent activation function.

The atom-environment descriptors are grouped into radial and angular categories. Each radial descriptor $q_n$ is formed by summing a radial function (indexed by $n$) over the neighbors; each angular descriptor $q_{nl}$ is formed by summing the product of a radial function and a spherical harmonic function (indexed by $l$ and $m$), followed by contraction over $m$, to make it rotational invariant. The radial functions themselves are linear combinations of a set of basis functions constructed via Chebyshev polynomials of the first kind, with trainable expansion coefficients. The radial functions in $q_n$ typically have a larger cutoff radius than those in $q_{nl}$. This difference allows the description of longer-range interactions with reduced computational cost compared to using the same cutoff radius for both $q_n$ and $q_{nl}$. In the original implementation [8], the neural network and descriptor parameters were optimized using a natural evolution strategy [22], which gives rise to the name NEP; however, gradient based alternative training algorithms are available in MatPL.

The high computational efficiency of NEP stems from all its design features, including the lightweight neural network architecture, the efficient iterative calculation of the Chebyshev polynomials, and the use of a shorter cutoff radius for the angular descriptors. The NEP approach as implemented in the GPUMD package [23] has demonstrated competitive computational efficiency on platforms with one to a few NVIDIA GPUs [24]. In this work, we port the NEP approach to LAMMPS and further optimize it for CNIS, expand the number of GPGPUs to tens of thousands. Although NEP is light weight compared to other MLFFs, its accuracy is on par with other more complex MLFFs especially for a given physics problem, as will be discussed in Section VII (Fig. 3). Using MatPL package, one can relatively quickly (1-3 weeks) develop a specific NEP potential for a given physical system, starting from a pretrained foundational MLFF model, and a small number of additional DFT calculations.

## IV. CURRENT STATE OF THE ART

For over four decades, *ab initio* molecular dynamics (AIMD) based on density functional theory (DFT) has been the gold standard for simulating atomic interactions with predictive capability. However, its inherent cubic scaling with system size has historically restricted its applications to systems of only thousands of atoms over picosecond timescales. The advance of MLFF has fundamentally altered this landscape, enabling simulations that combine the quantum mechanical accuracy with the computational efficiency of empirical force fields.

Table 1. Performance of molecular dynamics simulators with *ab initio* accuracy. The abbreviation TtS, DP and NEP stand for time-to-solution, Behler-Parrinello scheme, Deep Potential and Neuroevolution potential, respectively. The time-step of water system is 0.5 fs, and that of other systems is 1 fs.

| Application | Year | Potential | System | #atoms | #CPU Cores | #GPUs | Machine | Peak [FLOPS] | TtS [s/atom/step] |
|---|---|---|---|---|---|---|---|---|---|
| DeePMD-kit[16] | 2020 | DP | Cu | 127M | 27.3K | 27.3K | Summit | 91P | $8.1 \times 10^{-10}$ |
| SNAP ML-IAP[18] | 2021 | SNAP | C | 1B | 27.3K | 27.3K | Summit | 50P | $3.4 \times 10^{-11}$ |
| DeePMD-kit[25] | 2022 | DP | Cu | 3.4B | 27.3K | 27.3K | Summit | 43.7P | $1.1 \times 10^{-10}$ |
| Allegro[17] | 2023 | Allegro | HIV capsid | 100M | 1280 | 5120 | Perlmutter | - | $7.9 \times 10^{-11}$ |
| DeePMD-kit[26] | 2024 | DP | $H_2O$ | 29B | 35M | - | new Sunway | 57.1P | $8.8 \times 10^{-11}$ |
| This work (MIX16) | 2026 | NEP | $H_2O$ | 1.21T | 45K | 45K | CNIS | 246.0P | $7.7 \times 10^{-13}$ |
| This work (MIX16) | 2026 | NEP | Cu | 1.62T | 45K | 45K | CNIS | 271.2P | $5.0 \times 10^{-13}$ |
| This work (MIX16) | 2026 | NEP | $HfO_2$ | 0.93T | 45K | 45K | CNIS | 223.2P | $1.0 \times 10^{-12}$ |

Among these methods, the Deep Potential (DP) model has emerged as a leading approach, demonstrated to achieve DFT-level precision across a wide range of materials. The open-source DeePMD-kit package, integrated with the LAMMPS MD engine, has been the primary vehicle for deploying these models at scale. Early breakthroughs were realized on the Summit supercomputer, where optimized GPU kernels, mixed-precision arithmetic, and data-layout transformations enabled the first simulation of over 100 million atoms. Specifically, a 2020 implementation achieved 91 PFLOPS in double precision for a 127-million-atom copper system, achieving a time-to-solution of $8.1\times10^{-10}$ seconds per atom per step (s/atom/step)— a leap of more than three orders of magnitude over prior AIMD methods. Subsequent scaling efforts on Summit further pushed the spatial limit, with a 2022 study simulating 3.4 billion atoms of copper, albeit with a lower per-atom performance of $1.1\times10^{-10}$ s/atom/step due to increased communication overhead and memory constraints. Note, the large total number of atoms requires bigger memory, while higher time-to-solution (s/atom/step) performance requires efficient computation.

In the meantime, other MLFFs also achieved large scale MD simulations. The SNAP ML-IAP package demonstrated exceptional time-to-solution on Summit for a 1-billion-atom carbon system ($3.4\times10^{-11}$ s/atom/step), leveraging high arithmetic intensity and data-layout optimizations for short-range interactions. The Allegro model introduced a highly parallel, equivariant graph neural network architecture capable of scaling to 100 million atoms on the Perlmutter system, achieving a time-to-solution of $7.9\times10^{-11}$ s/atom/step for biomolecular systems. A 2024 port of the DeePMD-kit to the new Sunway supercomputer marked a significant milestone in absolute system size, scaling to 29 billion atoms for a water system on 35 million cores. This work introduced specialized parallelization and mixed-precision strategies tailored to the unique many-core architecture, attaining a peak performance of 57.1 PFLOPS and a time-to-solution of $8.8\times10^{-11}$ s/atom/step.

Despite these advances, the current state-of-the-art faces fundamental limitations in simultaneously achieving both extreme system size and sustained time-to-solution performance. As summarized in the Table 1, no prior demonstration has successfully combined the following three attributes: (1) a model with *ab initio* accuracy capable of capturing complex bonding and chemical heterogeneity, (2) a system size exceeding one trillion atoms, and (3) a time-to-solution below $10^{-11}$ seconds per atom per step. Previous works are either constrained by the memory hierarchy of GPUs, or the scalability of the underlying MLFF neural network architecture, or require massive CPU core counts that can compromise strong-scaling efficiency.

The current work advances the state of the art by achieving the first trillion-atom MD simulations at first-principles (DFT-level) accuracy. Using the NEP potential— a highly parallelizable, polynomial-based MLFF integrated with the LAMMPS GPU interface, with implementation improvements — we demonstrate highly efficient weak and strong scaling on CNIS. This represents a ~50× increase in system size over the largest prior DeePMD-kit result (29B atoms on new Sunway) and 100×, 1000× increase in the time-to-solution s/atom/step over the 29B atoms work and 2020 DeePMD work respectively, while preserving *ab initio* fidelity.

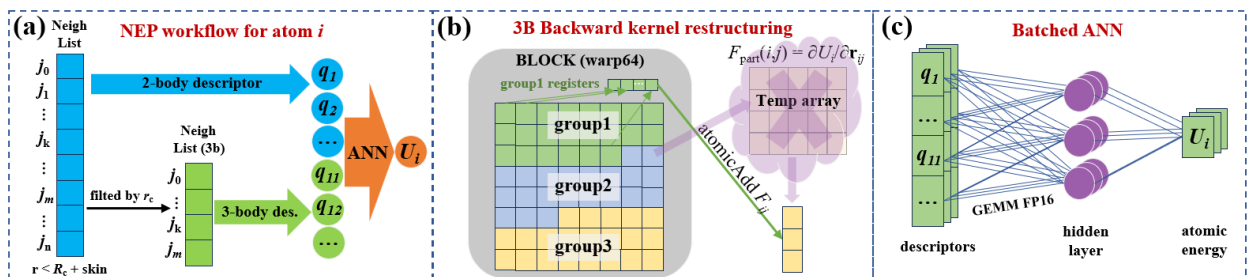

Fig. 2 Schematic overview of the key algorithmic innovations in the optimized NEP evaluation pipeline. (a) Overall computational workflow per MD step. For each atom, the two-body radial descriptors ($q_1, q_2, …$) and three-body angular descriptors ($q_{11}, q_{12}, …$) are computed from the neighbor list, followed by ANN inference to obtain the atomic energy $U_i$. The three-body neighbor list is filtered by a shorter cutoff $r_c$. (b) 3B Backward kernel restructuring. Neighbors of multiple center atoms (1-3 groups depend on the average number of angular neighbors) are processed within a warp-64 block, with all working arrays placed in registers. The partial forces are accumulated directly via atomicAdd, eliminating the temporary force buffer (crossed out). (c) Illustration of the batched ANN inference using FP16 GEMM on tensor cores. Atoms are processed collectively, with descriptors (blue, green), hidden-layer neurons (gray), and atomic energies $U_i$ (orange) forming matrix-multiplication operations for high throughput.

## V. INNOVATIONS REALIZED

### A. Summary of Contributions

Our primary contribution is the first-ever demonstration of trillion-atom molecular dynamics simulations with first-principles accuracy. This milestone is enabled by five hardware-aware algorithmic innovations that fully exploit the architectural features of the SIMT-based GPGPUs in the cluster—specifically their large on-chip register files (768 KB), limited shared memory (64 KB) and L2 cache (8 MB)—to accelerate the five-stage NEP evaluation pipeline. The innovations comprise: (1) kernel restructuring through compile-time template instantiation for register-centric data placement, together with artificial neural network (ANN) inference restructured as batched GEMM to exploit tensor-core units, and consolidation of the two-pass three-body backward into a single kernel, eliminating the large intermediate force buffer and its associated off-chip traffic; (2) geometry reuse across angular channels in the three-body descriptor via loop tiling, removing redundant bond-distance and spherical-harmonic evaluations; (3) an adaptive centers-per-block (CPB) scheduling policy that improves SIMD lane utilization in the three-body backward kernel by assigning multiple center atoms to a single thread block when the per-atom angular neighbor count is low, selected automatically based on the 90th-percentile neighbor count; and (4) selective half-precision (FP16) storage of the per-atom descriptor $q$, ANN gradients Fp, and angular accumulator sum_fxyz to relieve memory-bandwidth pressure and proportionally increase the maximum computable system size within a given GPU memory budget. Collectively, these optimizations achieve more than a 10× end-to-end speedup on a single GPU compared with the baseline implementation.

### B. Algorithmic Innovations

As illustrated in Fig. 2a, the NEP evaluation consists of five sequential stages per MD step: two-body radial descriptor, three-body angular descriptor, ANN inference with gradient computation, and two-body/three-body backward force accumulation. Our above 4 optimization strategies exploit three hardware characteristics of the SIMT-based GPGPU in the cluster that have major impacts on performance: a large on-chip register file (768 KB), limited shared memory (64 KB) and L2 cache (8 MB).

Table 2 reports the runtime of each NEP evaluation kernel for the $H_2O$ test case on a single GPU, broken down by the optimization techniques listed in the first column.

Table 2. Single-GPU ablation study on $H_2O$ (N = $6.75 \times 10^6$ atoms, 20 steps). The first column describes the optimization techniques which are described in Section V, algorithm innovations sub-section. Baseline I is the unmodified main-branch baseline code with kernel execution times measured via runtime API events; all other rows apply the optimization techniques described in Column 1 and report times by the same method. Row II bundles the three changes of Optimization 1 simultaneously—compile-time template instantiation for register placement, batched GEMM ANN inference, and three-body backward kernel consolidation—to isolate their combined effect before further tuning. All time units are in ms.

| Optimization techniques | 2B Descriptor | 3B Descriptor | ANN | 2B Backward | 3B Backward | Total (ms) | Speedup |
|---|---|---|---|---|---|---|---|
| I Baseline | 156 | 380.2 | — | 308.7 | 772.6 | 1917.8 | 1.0× |
| II Kernel template specialization + GEMM ANN + 3B Bwd kernel consolidation | 25.1 | 40.2 | 6.6 | 45.2 | 146.4 | 288.1 | 6.66× |
| III 3B descriptor channel loop tiling | 25.1 | 16.3 | 6.6 | 45.2 | 146.4 | 264 | 7.27× |
| IV Adaptive CPB scheduling for 3B Bwd. | 25.1 | 16.3 | 6.6 | 45.2 | 78.4 | 197.7 | 9.70× |
| V FP16 storage for q and Fp (bandwidth opt.) | 25 | 16.3 | 5.6 | 37.1 | 78.5 | 185.6 | 10.33× |
| VI FP16 storage for sum_fxyz (scale extension) | 25 | 15 | 5.6 | 37.1 | 78.8 | 181.9 | 10.54× |

Optimization 1. Kernel Restructuring: Registers and Batched ANN and atomic add (II in Table 2).

The baseline kernels allocate per-thread working arrays at runtime, placing them in shared memory or—when shared memory pressure is high—in slow off-chip memory, because the model shape is unknown at compile time. We address this by compiling a dedicated kernel variant for each supported model configuration, fixing all array sizes at compile time; the compiler can then map the per-thread working data onto the accelerator's large on-chip register file, as shown in Fig. 2b, substantially reducing traffic to slower memory tiers. Simultaneously, we restructure ANN inference from a per-atom serial loop into a batched matrix multiplication over all atoms in a block, enabling the use of the accelerator's high-throughput tensor-core units, as shown in Fig. 2c. We also consolidate the two-kernel three-body backward pipeline into a single kernel. The original two-pass design—staging partial forces into a large temporary array and scattering them in a second pass—was a common software strategy to avoid write conflicts across center atoms; on modern accelerators with hardware atomic units, however, direct accumulation via atomic operations is faster, a conclusion also reached by the broader MD community. Eliminating the temporary array (as shown in Fig. 2b) removes its read/write traffic and reduces peak memory footprint. Together these changes account for the 6.66× jump in row I→II of Table 2.

Profiling after this optimization reveals that the ANN kernel is now purely memory-bandwidth limited: the per-atom descriptor and gradient data reach hundreds of MB at simulation scale and overwhelm the 8 MB on-chip cache in every kernel call, while the network weights (a few KB) are reused freely. Measured FP32 tensor-core utilization is only 5–23% of the hardware tensor-core peak across the three test systems, confirming that the ANN kernel is bandwidth-bound rather than compute-bound. This finding directly motivates Optimization 4 to be described later.

Optimization 2. Three-Body Descriptor: Geometry Reuse Across Angular Channels (III in Table 2)

The angular descriptor for atom $i$ builds $N_{\max}^{\text{ang}}$ independent channel summed over all neighboring atoms $j$, as shown in line 3 of Algorithm 1.

| Algorithm 1: Tiled Geometry Reuse Across Angular Channels in Three-Body Descriptor |   |
|---|---|
| 1 | for n = 0; n < $N_{\max}^{\text{ang}}$; n += NTILE do |
| 2 |    s_buf[NTILE][24] ← 0        # with 3-body L$_{\max}$ = 4 |
| 3 |    for each neighbor j ∈ N_angular(i) do |
| 4 |       fc ← $f_c(r_{ij})$ |
| 5 |       fn12 ← radial_basis_functions($1/r_{ij}$, $r_{ij}$, fc) |
| 6 |       $Y_l^m$ ←spherical_harmonic ($x_{ij}, y_{ij}, z_{ij}, r_{ij}$) |
| 7 |       for t = 0 to NTILE-1 do #pragma unroll |
| 8 |          $g_n$ ← 0 |
| 9 |          for k = 0 to BASIS_SIZE_ANGULAR do     #pragma unroll |
| 10 |             $g_n$← fma(fn12[k], $C_{nk}^{IJ}$, $g_n$) |
| 11 |          accumulate_s_from_basis($g_n$, $Y_l^m$, s_buf[t]) |
| 12 |    for t = 0 to NTILE-1 do #pragma unroll |
| 13 |       store s_buf[t] into $q_{nl}$ |
| 14 |       store s_buf[t] into sum_fxyz for later angular force |

The radial weight $g_n$ (in line 10) differs for every channel $n$. The straightforward implementation sweeps over all neighbors j once per channel n—repeating the same bond-distance and spherical-harmonic evaluation $N_{\max}^{\text{ang}}$ times per neighbor j. In other words, the channel n is used as an outer loop with j being the inner loop. The key insight is that although the weights change with $n$, the geometry-dependent computations—specifically, the cutoff function $f_c(r_{ij})$ and the spherical harmonics $Y_l^m(x_{ij}, y_{ij}, z_{ij}, r_{ij})$ do not change. We therefore tile the outer channel loop with a compile-time tile size NTILE, as shown in line 1 of Algorithm 1: within each tile, all NTILE accumulators are updated during a single neighbor pass, so geometry is computed once and reused across NTILE consecutive channels, with the register file holding all intermediate values. The logical loop order ($n$ outer, $j$ inner) is preserved; only the granularity of the outer loop changes. For the model configurations tested here NTILE=5 is optimal. This reduces the 3B descriptor time from 40.2 ms to 16.3 ms (2.5×, row II→III) in Table 2 and yields a 1.09× end-to-end gain on the test systems, as the 3B backward stage remains the bottleneck at this point.

Optimization 3. Three-Body Backward: Adaptive Thread Utilization (IV in Table 1)

The kernel consolidation in Optimization 1 resolves the off-chip traffic problem of the three-body backward; the remaining bottleneck is SIMD lane utilization. In the baseline kernel, each thread block is assigned to exactly one center atom, so the block width of 64 threads must cover that atom's angular neighbors alone; when the neighbor count is well below 64—H$_2$O averages 31 angular neighbors—nearly half the lanes are idle per cycle. We introduce an adaptive *centers-per-block* (CPB) policy that assigns 1, 2, or 3 center atoms to a single thread block, automatically chosen based on the 90th-percentile neighbor count p$_{90}$:

$$\text{CPB} = \begin{cases} 3, & p_{90} \leq 20 \\ 2, & p_{90} < 48 \\ 1, & \text{others} \end{cases}$$

Here p$_{90}$ is the value where 90% of the cases that the number of neighbor atoms is below this number. No user tuning is required: the policy self-selects CPB = 3 for sparse metallic systems (Cu, p$_{90}$ = 18), CPB=2 for molecular fluids (H$_2$O, p$_{90}$ = 22) and ionic solids (HfO$_2$, p$_{90}$ = 22), and CPB=1 for dense, high-coordination systems. The Fig. 2b illustrates a representative case with CPB = 3, where the 64 threads within a warp are partitioned into three groups of 22, 21, and 21 threads, respectively, with each group responsible for processing one of the three central atoms. Because the 3B backward is the dominant bottleneck at this stage, this optimization delivers the largest single-step gain in the ablation: 1.34× end-to-end for the test H$_2$O system (row III→IV).

Optimization 4. Half-Precision (FP16) Storage to Relieve Bandwidth and Extend Scale (V, VI in Table 1)

Motivated by the bandwidth analysis in Optimization 1, we store three intermediate tensors—the per-atom descriptor $q$, the ANN gradient $F_p$, and the angular accumulator sum_fxyz—in half precision in GPU memory, as shown in Fig. 2c. With one exception, all arithmetic remains in FP32 regardless of storage format. The exception is the ANN forward pass: its tensor-core GEMM consumes $q$ directly as FP16 input, so the matrix multiply itself executes in FP16 precision; all subsequent ANN operations (activation, gradient) remain in FP32. As we will show in section VII, the loss in accuracy due to this FP16 usage can be ignored.

The per-stage gains are consistent with each tensor's role (row IV→V→VI of Table 2): storing $F_p$ in FP16 halves the read traffic of both backward kernels, with the largest visible effect on the 2B backward (45.2→37.1 ms); storing $q$ in FP16 reduces ANN input bandwidth (ANN: 6.6→5.6 ms); storing sum_fxyz in FP16 gives a modest descriptor speedup (16.3→15.0 ms) but little change in the 3B backward, whose bottleneck is per-neighbor arithmetic rather than tensor reads.

Besides running time, the FP16 sum_fxyz storage is a key factor for our ability of large-scale simulations: as the largest intermediate tensor in the pipeline, halving its footprint proportionally increases the maximum atom count achievable within a fixed GPU memory budget—a critical factor for our trillion-atom simulations.

## VI. HOW PERFORMANCE WAS MEASURED

### A. Physical Systems

To evaluate the computational efficiency and accuracy of our NEP model, we selected three distinct condensed-phase systems—water (insulating), copper (metallic), and hafnium dioxide ($HfO_2$, high-k dielectric)—covering a broad spectrum of chemical bonding and material properties. For each system, we tailored the NEP hyperparameters balancing the accuracy and efficiency: radial/angular cutoffs of 6.0/4.0 Å for water and Cu, and 8.0/4.0 Å for $HfO_2$. The neural network architecture consisted of a single hidden layer with 100 neurons for water, while 30 neurons are enough for both Cu and $HfO_2$. All models share a radial and angular basis of $n_{max} = 4$, a basis size of 8, three-body $L_{max} = 4$, and four-body $L_{max} = 2$. Short-range repulsive interactions were introduced via the Ziegler-Biersack-Littmark (ZBL) [27] screened potential for Cu (outer cutoff 2.0 Å) and $HfO_2$ (outer cutoff 1.5 Å), whereas water was simulated without ZBL. The NEP machine-learning potential trained via MatPL demonstrates excellent accuracy in reproducing DFT reference data. As shown in Fig. 3 for $H_2O$ system, the model achieves a root-mean-square error (RMSE) of 64.6 meV/Å for atomic forces. The parity plot exhibits a tight distribution around the ideal y = x line, and the error distribution histogram indicates that over 49% of the force predictions have absolute errors below 0.05 eV/Å, with the majority of errors concentrated in the low-error regime. For the per-atom virial (stress), the model yields an even higher accuracy with an RMSE of 4.5 meV/atom and a near-perfect $R^2$=1.000. Similarly, the potential energy per atom is predicted with an RMSE of 1.8 meV/atom and $R^2 = 1.000$. In both cases, the error distributions are strongly skewed toward zero, confirming that the vast majority of predictions lie within a few meV/atom of the DFT reference values. Molecular dynamics (MD) runs were carried out for 500 steps, requiring 501 force/energy evaluations. Timesteps were set to 0.5 fs for water and 1.0 fs for the other two systems. Initial atomic velocities were drawn from a Maxwell-Boltzmann distribution at 300 K. A neighbor-list buffer of 0.3 Å was employed and refreshed every 10 integration steps (check flag enabled). Thermodynamic observables—kinetic and potential energies, temperature, pressure—were recorded at intervals of 100 steps. This setup enables a rigorous and consistent performance assessment of our NEP potentials across insulating, metallic, and mixed ionic-covalent regimes.

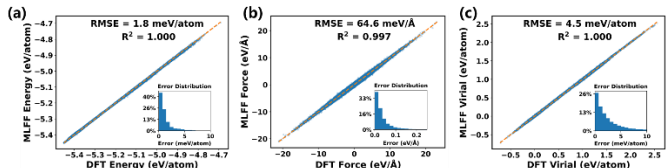

Fig 3. Training results of NEP model for $H_2O$. (a)-(c) The comparison between the MatPL/NEP predictions and DFT values of energy, force and virial.

We performed scaling measurements across all three target systems—water, copper, and hafnium dioxide—to assess the parallel efficiency of our NEP-based MD simulations. For water, strong scaling was evaluated on a system containing 110,578,630,656 atoms. In weak-scaling mode, the per-GPU atom count was fixed at 26,996,736; scaling up to 45,000 GPUs yielded a global system exceeding one trillion atoms. For copper, the strong-scaling system comprised 147,438,174,208 atoms. Its weak-scaling baseline was set to 35,995,648 atoms per GPU, with the largest configuration reaching 1,619,804,160,000 atoms across 45,000 GPUs. For $HfO_2$, the strong-scaling test employed a system of 84,934,656,000 atoms, while weak scaling started from 20,736,000 atoms per GPU.

### B. HPC Platform and Software Environment

All performance measurements were conducted on CNIS. The system employs a heterogeneous cluster architecture combining scale-up node design with a scale-out inter-connection fabric. Each compute node integrates two 64-bit CISC-based server processors and eight SIMT-based GPGPU computing accelerators. The node processor is a 64-core, 2.4 GHz CISC-based chip supporting a NUMA memory topology. The memory subsystem comprises eight channels of DDR5 running at 6400 MT/s. The I/O complex provides PCIe Gen5 connectivity to the attached accelerators, sustaining a bidirectional host-to-device and device-to-host bandwidth of 64 GB/s per accelerator link. Each accelerator is a SIMT-based GPGPU device containing 320 SIMD units. The accelerator is equipped with 64 GB of on-package High Bandwidth Memory (HBM), which delivers a peak memory bandwidth of 1.8 TB/s. The on-chip memory hierarchy includes a 768 KB register file, 64 KB of Local Data Share (LDS), and an 8 MB L2 cache. The single-GPU theoretical peak FP64 performance is 32.7 TFLOPS. Inter-node communication is provided by a proprietary, InfiniBand-like native RDMA interconnect. Each compute node is equipped with four network ports, each capable of 400 Gbps line rate. The system-wide network topology is organized as a three-tier Clos fabric with two independent data planes, offering high bisection bandwidth and robust performance for both point-to-point and collective operations.

Table 3. Software environment

| Component | Specification / Version |
|---|---|
| operating System | Anolis OS 8.9 |
| Host compiler | GCC 8.5.0 |
| Accelerator Program | compatible with mainstream GPGPU API standards |
| MPI | OpenMPI 5.0.3 |
| Collective Communication Library | Collective communications library compatible with RCCL APIs |

The software environment used for all performance measurements is summarized in Table 3. The operating system is Anolis OS 8.9. The accelerator software stack is based on a GPGPU programming environment compatible with mainstream GPGPU API standards, ensuring broad portability. The CPU toolchain relies on GCC 8.5.0, BLIS 2.0, and FFTW 3.3.10. Distributed-memory parallelism is managed via OpenMPI 5.0.3, complemented by a collective communications library compatible with RCCL APIs.

*C. Measurement Methodology*

The current CNIS lacks a reliable operation counter. All floating-point operation (FLOP) counts reported in this work are derived by static instruction-level analysis of the GPU kernel source code. For each computational stage we enumerate every floating-point arithmetic instruction executed per atom-pair or per atom interaction, multiply by the corresponding interaction count measured at runtime, and sum across the full MD step. To establish that the static-analysis approach yields trustworthy results, we performed the cross-validation described below.

We compiled the unmodified upstream branch of the NEP GPU code with CUDA and profiled it on NVIDIA GPU using NVIDIA Nsight Compute (NCU). NCU reports per-kernel FP32 and FP64 instruction counts, which give a ground-truth FLOP measurement independent of any analytical model. We then applied our static-analysis method to the corresponding kernels of our code and compared the results for three representative systems (Cu, $H_2O$, $HfO_2$) across four computational stages: two-body (2B) descriptor construction, combined three-body (3B) descriptor + neural-network (ANN) evaluation, two-body (2B) force back-propagation, and three-body (3B) force back-propagation.

All results are averaged over 11 consecutive MD steps to suppress step-to-step variance. As shown in Table 4, the estimated values agree closely with NCU measurements across all five computational stages for Cu, water, and $HfO_2$ systems, with relative differences of less than 4% in total kernel time and below 10% for individual stages for all three systems. This confirms that our instrumentation-based FLOP counting method provides a reliable and slightly conservative estimate of sustained performance, even on the target GPU architecture with limited hardware counter support.

Table 4. Cross-validation of FLOP count. "NCU" = FP32 + FP64 hardware counter sum reported by Nsight Compute for the main-branch kernel(s) corresponding to each stage. "Estimate" = static-analysis result for each stage using our counting methodology applied to the optimized implementation. "diff" = (NCU − Estimate) / Estimate. Units: GFLOPs per MD step.

| Stage | Cu | | | Water | | | $HfO_2$ | | |
|---|---|---|---|---|---|---|---|---|---|
| | NCU | Estimate | diff | NCU | Estimate | diff | NCU | Estimate | diff |
| 2B Descriptors | 0.728 | 0.668 | 9.00% | 2.903 | 2.662 | 9.10% | 3.126 | 2.907 | 7.50% |
| 3B Desc + ANN | 1.421 | 1.465 | −3.0% | 9.000 | 8.901 | 1.10% | 2.825 | 2.903 | −2.7% |
| 2B Backward | 1.856 | 1.684 | 10.20% | 7.392 | 6.738 | 9.70% | 7.925 | 7.291 | 8.70% |
| 3B Backward | 4.167 | 4.157 | 0.20% | 21.418 | 21.365 | 0.20% | 8.576 | 8.554 | 0.30% |
| Total | 8.172 | 7.974 | 2.50% | 40.713 | 39.666 | 2.60% | 22.452 | 21.655 | 3.70% |

. We report two important performance metrics:
- **Time-to-solution**: defined as $\frac{\text{MD loop time}}{\text{number of atoms} \times \text{number of MD steps}}$, with a unit of s/atom/step. The "MD loop time" is the wall clock time used in the MD loop for 500 MD steps, including some limited IO like the total energy reports. Setup time, such as the initial setup of the system and MPI initialization and finalization, are not included.
- **Sustained performance**: defined as $\frac{\text{total FLOPs}}{\text{MD loop time}}$ with a unit of PFLOPS. The total FLOPs count is provided by the above static instruction-level analysis method. In Table.1, we report it in the Peak column.

## VII. PERFORMANCE RESULTS

*A. Single-GPU Performance and Mixed-Precision Validation*

1. Single-GPU Performance

This optimization was evaluated on representative local systems containing several million atoms. As summarized in Table 5, the optimized FP32 kernel delivers dramatic speedups compared to the original implementation: 9.67× for FCC Cu (4.50 million atoms), 9.70× for liquid water (6.75 million atoms), and 6.63× for $HfO_2$ (5.18 million atoms).

In terms of absolute performance, the Matom-step/s metric (the inverse of the time-to-solution) improved by nearly the same factors — from 4.78 to 46.19 (9.67×) for Cu, 3.52 to 34.14 (9.70×) for water, and 3.51 to 23.24 (6.63×) for $HfO_2$.

These optimizations significantly lower the per-step computational cost while maintaining the numerical stability and accuracy required for machine-learning potentials. By accelerating the most time-consuming kernels without sacrificing portability or correctness, the enhanced FP32 pathway contributes directly to the excellent weak- and strong-scaling performance observed at trillion-atom scales, enabling higher throughput and shorter time-to-solution on large-scale GPU supercomputers.

Table 5. Performance of optimized FP32 kernel version for the time of each MD step on one GPU.

| system | number of atoms | baseline [ms/step] | Opt FP32 [ms/step] | speedup |
|---|---|---|---|---|
| Cu | 4,499,456 | 941.4 | 97.4 | 9.67× |
| water | 6,749,184 | 1917.8 | 197.7 | 9.70× |
| HfO$_2$ | 5,184,000 | 1478.8 | 223.1 | 6.63× |

2. Mixed-Precision Validation

As we show in section V, we selectively store three intermediate tensors (descriptor, ANN gradient and angular accumulator) in FP16 to reduce off-chip memory traffic, while keeping all arithmetic in FP32; the sole exception is the ANN tensor-core GEMM, which consumes FP16 descriptor input natively. To quantify the impact of optimized mixed precision on simulation accuracy, we evaluated the NEP model using two mixed-precision configurations: MIX-32 (FP32 accumulation with FP64 critical operations) and MIX-16 (lower-precision mixed mode). Taking the double-precision NEP model (whose predictions show excellent agreement with DFT references, as presented in Fig. 3) as the ground-truth baseline, we computed the test errors on an independent validation set consisting of 100 water configurations, each containing 216 water molecules (the error will not aggregate with number of atoms since the NEP has a local descriptor formalism).

Table 6. Test errors for the water system from models with different precision

| Precision | Error in energy [eV/molecule] | Error in Force [eV/Å] |
|---|---|---|
| MIX-32 | 7.27×10$^{-6}$ | 4.38×10$^{-6}$ |
| MIX-16 | 5.73×10$^{-5}$ | 4.59×10$^{-5}$ |

As summarized in Table 6, the MIX-32 configuration achieves extremely low errors, with a mean absolute error of 7.27×10$^{-6}$ eV/molecule in potential energy and 4.38×10$^{-6}$ eV/Å in atomic forces. Even with the more aggressive MIX-16 mode, the errors remain remarkably small — 5.73×10$^{-5}$ eV/molecule for energy and 4.59×10$^{-5}$ eV/Å for forces. These results confirm that mixed precision introduces negligible error while unlocking extreme performance, both far lower than the NEP error compared with DFT.

The minimal loss in numerical fidelity when moving from MIX-32 to MIX-16 enables significant performance gains without compromising the physical reliability of the simulations. This careful balance between precision and speed is essential for achieving both high accuracy and extreme scalability on modern GPU architectures, allowing trillion-atom molecular dynamics simulations to be performed at sustained high throughput while retaining near-DFT-level fidelity.

*B. Weak Scaling*

We performed weak-scaling tests on three representative systems: liquid water, cubic HfO$_2$, and FCC Cu. In these benchmarks, the number of atoms was increased proportionally with the number of GPUs to maintain roughly constant workload per GPU. The tests span from 1 GPU (system sizes of approximately 2–3.6×10$^7$ atoms) to 45,000 GPUs (system sizes reaching 9.3×10$^{11}$ to 1.62×10$^{12}$ atoms), covering more than four orders of magnitude in both processor count and problem size.

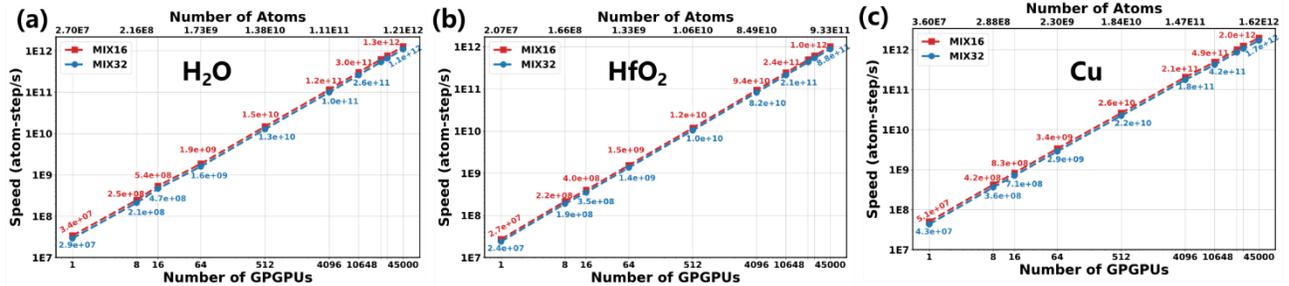

Fig. 4 Weak scaling: (a) the water system. Number of atom ranges from 26,996,736 to 1,214,853,120,000. (b) the HfO$_2$ system. Number of atom ranges from 20,736,000 to 933,120,000,000. (c) the copper system. Number of atom ranges from 35,995,648 to 1,619,804,160,000. Number of GPU ranges from 1 to 45,000 in all three systems.

As shown in Fig. 4a (liquid water), the simulation speed scales nearly linearly with the number of GPUs. Using MIX16 precision, the speed increases from 3.4×10$^7$ atom-step/s on a single GPU (6.539 TFLOPS) to 1.3×10$^{12}$ atom-step/s on 45,000 GPUs. This corresponds to a time-to-solution of 7.69×10$^{-13}$ s/step/atom and peak performance of 246.0 PFLOPS. The MIX32 mode achieves 1.1×10$^{12}$ atom-step/s at the largest scale, corresponding to 9.09×10$^{-13}$ s/step/atom and 208.2 PFLOPS. Parallel efficiency remains excellent throughout, staying above 83% even at full system size. The MD loop time increases only modestly, reflecting highly effective overlap between computation and communication, and the problem is local computation dominated.

For cubic HfO$_2$ (Fig. 4b), which features more complex oxide chemistry, the implementation also exhibits outstanding weak scaling. MIX16 performance rises from 2.7×10$^7$ atom-step/s (1 GPU with 6.064 TFLOPS) to 1.0×10$^{12}$ atom-step/s (45,000 GPUs), yielding a time-to-solution of 1.00×10$^{-12}$ s/step/atom and 223.2 PFLOPS. MIX32 reaches 8.8×10$^{11}$ atom-step/s (1.14×10$^{-12}$ s/step/atom and 196.416 PFLOPS).

Efficiency remains robust at approximately 82% at the largest scale.

The FCC Cu system (Fig. 4c) delivers the highest overall performance, with MIX16 speed scaling from $5.1\times10^7$ atom-step/s on 1 GPU (6.934 TFLOPS) to $2.0\times10^{12}$ atom-step/s on 45,000 GPUs ($5\times10^{-13}$ s/step/atom and 271.2 PFLOPS) and MIX32 reaching $1.7\times10^{12}$ atom-step/s ($5.88\times10^{-13}$ s/step/atom and 230.5 PFLOPS). Efficiency stays above 86% at the maximum scale, with values close to 100% in several intermediate regimes due to favorable cache effects and optimized neighbor-list construction in metallic systems. This results in the highest sustained throughput observed across all tested systems.

Across the three systems, the MatPL NEP-LAMMPS interface achieves near-ideal weak scaling over four orders of magnitude in GPU count and system size, sustaining parallel efficiencies of 82–87% at 45,000 GPUs while pushing the problem size into the trillion-atom regime. Minor fluctuations in efficiency at the largest scales are primarily attributable to variations in network topology and collective communication patterns on the supercomputer. The consistent use of mixed-precision (MIX16 and MIX32) strategies enables high computational throughput without compromising the accuracy required for machine-learning potentials.

At the full scale of 45,000 GPUs, the MatPL NEP-LAMMPS interface delivers substantial sustained performance on trillion-atom systems. Based on single-GPU kernel floating point operation count analysis and the measured weak-scaling parallel efficiencies (83.6% for water, 86.9% Cu, and 81.8% for $HfO_2$), the estimated sustained performance reaches:

- 271.2 PFLOPS (MIX16) and 230.5 PFLOPS (MIX32) for FCC Cu (1.62 trillion atoms)
- 223.2 PFLOPS (MIX16) and 196.4 PFLOPS (MIX32) for cubic $HfO_2$ (0.933 trillion atoms)
- 246.0 PFLOPS (MIX16) and 208.2 PLOPS (MIX32) for liquid water (1.21 trillion atoms)

Compared to prior state-of-the-art MLFF-based MD implementations (see Table 1), our work achieves significantly larger system sizes — reaching 1.62 trillion atoms for Cu, 1.21 trillion atoms for water, and 0.933 trillion atoms for $HfO_2$ — while maintaining high parallel efficiencies of 82–87%. The largest previously reported MLFF run on 29 billion atoms (Table. 1). Our highest sustained MIX32 performance of 230.52 PFLOPS is about 15.7% of the FP64 peak theoretical performance of CNIS. Unlike some heavy computational tasks, such as the Quantum mechanical calculations, the light weight MD simulations traditionally have low sustained FLOPS performances. Nevertheless, it is 3 to 5 times larger than the previously reported FLOPS of MD simulations. The highest previous FLOPS performance was achieved due to the heavy computational need of deep neural network. But that is at the cost of worse time to solution result. That is why later works while improving the time to solution, their sustained performance went down. In terms of time to solution, our work at $5-10\times10^{-13}$ s/step/atom is $100\times$ better than the DeePMD-kit work of 2024, and $1000\times$ better than the 2020 work, and still our sustained performance is a few times higher than previously reported MD results.

### C. Strong Scaling

We evaluated the strong-scaling behavior of the MatPL NEP-LAMMPS interface on fixed, ultra-large systems: 111 billion atoms for liquid water, 84.9 billion atoms for cubic $HfO_2$, and 147 billion atoms for FCC Cu. However, the even bigger systems are not used since they cannot be run on smaller number GPUs. These tests were conducted from 4,096 GPUs to 45,000 GPUs to assess performance gains as the number of processors increases while the problem size remains constant.

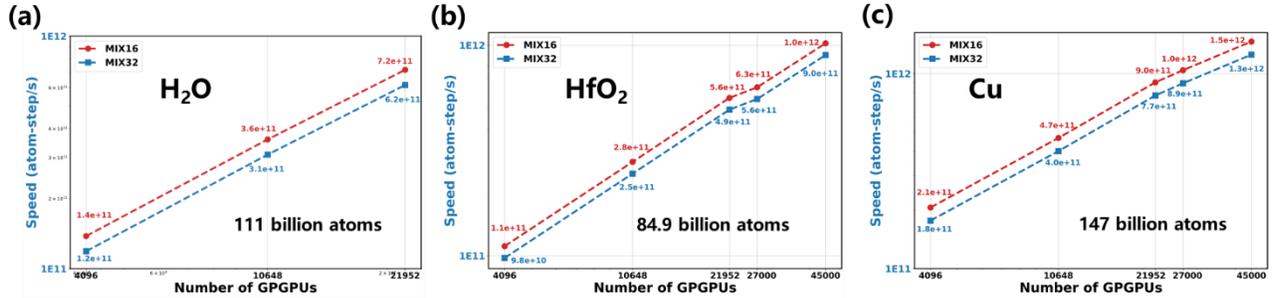

Fig 5: Strong scaling: (a) the water system of 110,578,630,656 atoms. (b) the $HfO_2$ system of 84,934,656,000 atoms. (c) the copper system of 147,438,174,208 atoms.

As shown in Fig. 5a (liquid water, 111 billion atoms), the simulation speed scales effectively with GPU count. Using MIX16 precision, speed increases from $1.4\times10^{11}$ atom-step/s at 4,096 GPUs to $7.2\times10^{11}$ atom-step/s at 21,952 GPUs, corresponding to a $5.1\times$ speedup (96.8% efficiency) and approximately 6.49 MD steps per second. The MIX32 mode reaches $6.2\times10^{12}$ atom-step/s at the same scale. The scaling remains nearly linear, indicating efficient workload distribution across thousands of additional GPUs.

For cubic $HfO_2$ (Fig. 5b, 84.9 billion atoms), strong scaling is exceptionally good. MIX16 performance rises from $1.1\times10^{11}$ atom-step/s on 4,096 GPUs to $1.0\times10^{12}$ atom-step/s on 45,000 GPUs — a $9.1\times$ improvement (83.7% efficiency) — enabling roughly 11.78 MD steps per second. MIX32 achieves $9.0\times10^{11}$ atom-step/s at full scale. The near-ideal linear trend persists to the maximum GPU count, demonstrating the interface's robustness for complex oxide systems even at extreme processor scales.

The FCC Cu system (Fig. 5c, 147 billion atoms) delivers the highest performance. MIX16 speed grows from $2.1\times10^{11}$ atom-step/s at 4,096 GPUs to $1.5\times10^{12}$ atom-step/s at 45,000 GPUs, achieving a $7.1\times$ speedup (64.6% efficiency) and approximately 10.20 MD steps per second. MIX32 reaches $1.3\times10^{12}$ atom-step/s at the largest scale. The scaling curve stays close to ideal linearity, benefiting from optimized neighbor-list construction and force evaluation in dense metallic systems.

## VIII. IMPLICATIONS

### A. Scientific Applications Enabled by Trillion-Atom Simulations

"Interface is the device" [28]—the performance of semiconductor devices is largely governed by interfacial structures. In advanced nodes, atomic-scale intermixing and dopant distribution at interfaces critically determine carrier transport and contact resistance [29–31]. Experimentally, the fabrication and characterization of surface and interface structures often rely on trial-and-error methods, which consume substantial quantities of high-purity materials, expensive equipment time, and manpower, while offering limited efficiency. Theoretical simulations enable rapid screening of material combinations, interface configurations, doping schemes, and defect designs, thereby focusing experimental efforts on the most promising candidates and significantly shortening the development cycle. However, realistic device structures typically contain tens to hundreds of millions of atoms, far exceeding the capability of first-principles methods. As a result, previous interfacial simulations of field-effect transistors (FETs) have been restricted to empirical force fields or continuum models [32, 33], which suffer from limited accuracy in capturing fine interfacial features. In this work, by employing the NEP MLFF and the parallel optimization capabilities of the MatPL on the CNIS machine, the atomic-scale interfacial simulations become feasible for realistic structures comprising multiple FET units, which may constitute a logic cell in the electric circuit design.

In Fig. 6, we present a molecular dynamics simulation result of interfacial interdiffusion in an FET interconnect system containing approximately 100 million atoms using the MatPL NEP-LAMMPS code. The simulation cell measures $300\times240\times372$ nm³, approaching realistic FET dimensions (Fig. 6c), and was constructed by expanding pre-annealed units with a consistent cross-sectional profile. Periodic boundary conditions and the NVT ensemble were employed, with the temperature maintained at 300 K. A time step of 1 fs was used, and the simulation ran for 50,000 steps. Here we used a smaller system instead of the full one trillion atom system, mainly to allow us to do a longer time run for metal diffusion, taking advantage of the high time-to-solution capability illustrated in the current work. The results reveal that the Au/Si interfaces at different locations consistently exhibit a distinct transition region with a thickness of approximately 1 nm, rather than forming ideal sharp interfaces. Specifically, the Si concentration gradually decreases from its intrinsic value on the silicon substrate side, yet exhibits a local enrichment peak within the transition region, ultimately approaching zero on the gold electrode side (Fig. 6b). These findings indicate that the atomic arrangement at the Au/Si interface in large-scale FET devices undergoes significant alteration, deviating considerably from previously assumed ideal models.

The implications for science and engineering are far-reaching. In semiconductor technology, realistic simulations of $HfO_2$ at the trillion-atom scale enable direct investigation of defect dynamics, oxygen vacancy migration, grain boundary effects, polarization switching, and device reliability under operational conditions for systems far bigger than single transistors, thus allows us to do logic cell level simulations. These insights can accelerate the development of next-generation ferroelectric transistors (FeFETs), 3D FeNAND memory, embedded nonvolatile memory, and low-power logic devices critical for neuromorphic computing and high-density storage. In energy technologies, trillion-atom simulations will allow us to study a small patch of cathode materials consisting of multiple cathode particles, binding polymer, liquid electrolytes, and conductor materials, thus a realistic "eco" system. Being able to simulate such complex multiscale system all on atomic level will significantly enhance our ability to understand the underlying mechanism of such systems.

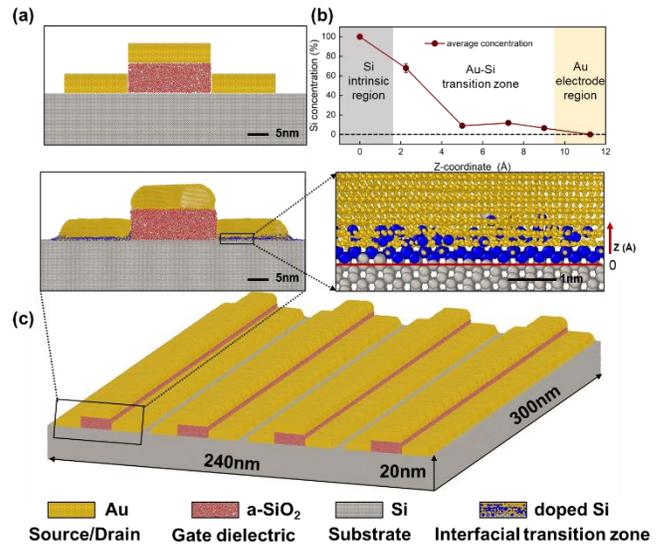

Fig. 6 Molecular dynamics simulation reveals interfacial atomic diffusion behavior and three-dimensional structural morphology in multi-channel FET devices. (a) Cross-sectional schematic of the initial FET unit structure (top) and cross-section of the FET unit after MD equilibration (bottom). Blue atoms mark Si atoms that have diffused into the Au lattice. Scale bar: 5 nm. (b) Silicon concentration profile along the Z-direction (top), quantitatively showing the average concentration gradient of Si atoms in the intrinsic Si region, the Au/Si interfacial transition region, and the Au electrode region. The lower panel presents a magnified view of the interfacial region; the red line marks the bottom of the transition region, defined as the concentration zero point. Scale bar: 1 nm. (c) Three-dimensional perspective view of the complete device structure. The dashed lines indicate the spatial correspondence between the local cross-sectional enlargement shown in (a) and the three-dimensional structure of the full device.

Beyond these domains, the capability of this work opens new frontiers in radiation damage assessment for nuclear materials, crack propagation and mechanical failure in amorphous and nanocrystalline alloys, geochemical and environmental processes at solid-liquid interfaces, and collective phenomena in biological and quantum materials. By bridging atomic-scale quantum mechanics with mesoscopic and experimental-relevant length and the fast time scale for smaller systems due to their good strong scaling, this work enables statistical sampling of rare events and long-time relaxation processes that govern real-world material performance, thereby transforming discovery pipelines in materials design, sustainable energy, and advanced manufacturing.

*B. Implications for Exascale Computing and Beyond*

For the broader HPC community, the most important implication lies in the generalizability of the solution. The neuroevolution potential (NEP) is tightly integrated with the widely used LAMMPS GPU interface. This makes extreme-scale, quantum-accurate simulations immediately accessible to thousands of researchers worldwide without proprietary development. Our work also demonstrates that even light weighted relatively modest arithmetic intensity jobs (like the light weighted NEP neural network) can achieve outstanding weak and strong scaling provided algorithmic innovation are done focused on memory efficiency and parallelization.

Looking forward, this work provides valuable lessons for future exascale and post-exascale systems: it highlights the importance of co-design between compact, highly parallelizable MLFF and GPU architectures and balance FLOPS performance and time-to-solution optimization and test the usage of lower precision operation on carefully selected steps, and shows how community-standard codes can be extended to unprecedented scales. As MLFF continue to evolve, the techniques developed here will help sustain performance gains on increasingly heterogeneous hardware while broadening access to transformative computational capabilities.

ACKNOWLEDGMENT

L. W. acknowledges the National Natural Science Foundation of China (Grant Nos. T2293700 and T2293702). F. D. acknowledges the National Science and Technology Major Project (2026ZD062500), the New Generation Artificial Intelligence-National Science and Technology Major Project (2025ZD0121802) and the Research Program from Suzhou Laboratory (SK-1502-2024-055), C. Q. acknowledges the China Postdoctoral Science Foundation (2025M781003). Z. F., P. G and Y. S. acknowledge Advanced Materials--National Science and Technology Major Project (No. 2024ZD0606900).

REFERENCES

[1] Hull, D. and Bacon, D.J. eds. 2011. *Introduction to dislocations*. Butterworth-Heinemann.
[2] Jung, S., Gwon, H., Hong, J., Park, K., Seo, D., Kim, H., Hyun, J., Yang, W. and Kang, K. 2014. Understanding the Degradation Mechanisms of $LiNi_{0.5}Co_{0.2}Mn_{0.3}O_2$ Cathode Material in Lithium Ion Batteries. *Advanced Energy Materials*. 4, 1 (Jan. 2014), 1300787. https://doi.org/10.1002/aenm.201300787.
[3] Li, J. and Huang, W. 2018. From Multiscale to Mesoscience: Addressing Mesoscales in Mesoregimes of Different Levels. *Annual Review of Chemical and Biomolecular Engineering*. 9, 1 (June 2018), 41–60. https://doi.org/10.1146/annurev-chembioeng-060817-084249.
[4] Behler, J. 2021. Four Generations of High-Dimensional Neural Network Potentials. *Chemical Reviews*. 121, 16 (Aug. 2021), 10037–10072. https://doi.org/10.1021/acs.chemrev.0c00868.
[5] Unke, O.T., Chmiela, S., Sauceda, H.E., Gastegger, M., Poltavsky, I., Schütt, K.T., Tkatchenko, A. and Müller, K.-R. 2021. Machine Learning Force Fields. *Chemical Reviews*. 121, 16 (Aug. 2021), 10142–10186. https://doi.org/10.1021/acs.chemrev.0c01111.
[6] Kocer, E., Ko, T.W. and Behler, J. 2022. Neural Network Potentials: A Concise Overview of Methods. *Annual Review of Physical Chemistry*. 73, 1 (Apr. 2022), 163–186. https://doi.org/10.1146/annurev-physchem-082720-034254.
[7] Xia, J., Zhang, Y. and Jiang, B. 2025. The evolution of machine learning potentials for molecules, reactions and materials. *Chemical Society Reviews*. 54, 10 (2025), 4790–4821. https://doi.org/10.1039/D5CS00104H.
[8] Fan, Z., Zeng, Z., Zhang, C., Wang, Y., Song, K., Dong, H., Chen, Y. and Ala-Nissila, T. 2021. Neuroevolution machine learning potentials: Combining high accuracy and low cost in atomistic simulations and application to heat transport. *Physical Review B*. 104, 10 (Sept. 2021), 104309. https://doi.org/10.1103/PhysRevB.104.104309.
[9] Suo, P., Wu, X., Tian, H. and Wang, L. 2026. Towards Scalable and Efficient Machine-Learning Force Fields: The MatPL package and Its Advancements on Neuroevolution Potentials. *ChemRxiv*. (2026). https://doi.org/https://doi.org/10.26434/chemrxiv.15001665.
[10] Thompson, A.P. et al. 2022. LAMMPS - a flexible simulation tool for particle-based materials modeling at the atomic, meso, and continuum scales. *Computer Physics Communications*. 271, (Feb. 2022), 108171. https://doi.org/10.1016/j.cpc.2021.108171.
[11] Behler, J. and Parrinello, M. 2007. Generalized Neural-Network Representation of High-Dimensional Potential-Energy Surfaces. *Physical Review Letters*. 98, 14 (Apr. 2007), 146401. https://doi.org/10.1103/PhysRevLett.98.146401.
[12] Bartók, A.P., Payne, M.C., Kondor, R. and Csányi, G. 2010. Gaussian Approximation Potentials: The Accuracy of Quantum Mechanics, without the Electrons. *Physical Review Letters*. 104, 13 (Apr. 2010), 136403. https://doi.org/10.1103/PhysRevLett.104.136403.
[13] Thompson, A.P., Swiler, L.P., Trott, C.R., Foiles, S.M. and Tucker, G.J. 2015. Spectral neighbor analysis method for automated generation of quantum-accurate interatomic potentials. *Journal of Computational Physics*. 285, (Mar. 2015), 316–330. https://doi.org/10.1016/j.jcp.2014.12.018.
[14] Wang, H., Zhang, L., Han, J. and E, W. 2018. DeePMD-kit: A deep learning package for many-body potential energy representation and molecular dynamics. *Computer Physics Communications*. 228, (July 2018), 178–184. https://doi.org/10.1016/j.cpc.2018.03.016.
[15] Musaelian, A., Batzner, S., Johansson, A., Sun, L., Owen, C.J., Kornbluth, M. and Kozinsky, B. 2023. Learning local equivariant representations for large-scale atomistic dynamics. *Nature Communications*. 14, 1 (Feb. 2023), 579. https://doi.org/10.1038/s41467-023-36329-y.
[16] Jia, W., Wang, H., Chen, M., Lu, D., Lin, L., Car, R., Weinan, E. and Zhang, L. 2020. Pushing the Limit of Molecular Dynamics with Ab Initio Accuracy to 100 Million Atoms with Machine Learning. *SC20: International Conference for High Performance Computing, Networking, Storage and Analysis* (Atlanta, GA, USA, Nov. 2020), 1–14.
[17] Kozinsky, B., Musaelian, A., Johansson, A. and Batzner, S. 2023. Scaling the Leading Accuracy of Deep Equivariant Models to Biomolecular Simulations of Realistic Size. *Proceedings of the International Conference for High Performance Computing, Networking, Storage and Analysis* (Denver CO USA, Nov. 2023), 1–12.
[18] Nguyen-Cong, K., Willman, J.T., Moore, S.G., Belonoshko, A.B., Gayatri, R., Weinberg, E., Wood, M.A., Thompson, A.P. and Oleynik, I.I. 2021. Billion atom molecular dynamics simulations of carbon at extreme conditions and experimental time and length scales. *Proceedings of the International Conference for High Performance Computing, Networking, Storage and Analysis* (St. Louis Missouri, Nov. 2021), 1–12.
[19] Liang, T. et al. 2025. NEP89: Universal neuroevolution potential for inorganic and organic materials across 89 elements. (Apr. 2025). https://doi.org/10.48550/arXiv.2504.21286.
[20] Dong, H., Shi, Y., Ying, P., Xu, K., Liang, T., Wang, Y., Zeng, Z., Wu, X., Zhou, W., Xiong, S., Chen, S. and Fan, Z. 2024. Molecular dynamics simulations of heat transport using machine-learned

[20] potentials: A mini-review and tutorial on GPUMD with neuroevolution potentials. *Journal of Applied Physics*. 135, 16 (Apr. 2024), 161101. https://doi.org/10.1063/5.0200833.

[21] Ying, P., Qian, C., Zhao, R., Wang, Y., Xu, K., Ding, F., Chen, S. and Fan, Z. 2025. Advances in modeling complex materials: The rise of neuroevolution potentials. *Chemical Physics Reviews*. 6, 1 (Mar. 2025), 011310. https://doi.org/10.1063/5.0259061.

[22] Schaul, T., Glasmachers, T. and Schmidhuber, J. 2011. High dimensions and heavy tails for natural evolution strategies. *Proceedings of the 13th annual conference on Genetic and evolutionary computation* (Dublin Ireland, July 2011), 845–852.

[23] Xu, K. et al. 2025. GPUMD 4.0: A high-performance molecular dynamics package for versatile materials simulations with machine-learned potentials. *Materials Genome Engineering Advances*. (Aug. 2025), e70028. https://doi.org/10.1002/mgea.70028.

[24] Song, K. et al. 2024. General-purpose machine-learned potential for 16 elemental metals and their alloys. *Nature Communications*. 15, 1 (Nov. 2024), 10208. https://doi.org/10.1038/s41467-024-54554-x.

[25] Guo, Z., Lu, D., Yan, Y., Hu, S., Liu, R., Tan, G., Sun, N., Jiang, W., Liu, L., Chen, Y., Zhang, L., Chen, M., Wang, H. and Jia, W. 2022. Extending the limit of molecular dynamics with *ab initio* accuracy to 10 billion atoms. *Proceedings of the 27th ACM SIGPLAN Symposium on Principles and Practice of Parallel Programming* (Seoul Republic of Korea, Apr. 2022), 205–218.

[26] Wang, X., Meng, X., Guo, Z., Li, M., Liu, L., Li, M., Xiao, Q., Zhao, T., Sun, N., Tan, G. and Jia, W. 2025. 29-Billion Atoms Molecular Dynamics Simulation With Ab Initio Accuracy on 35 Million Cores of New Sunway Supercomputer. *IEEE Transactions on Computers*. 74, 5 (May 2025), 1634–1648. https://doi.org/10.1109/TC.2025.3540646.

[27] Liu, J., Byggmästar, J., Fan, Z., Qian, P. and Su, Y. 2023. Large-scale machine-learning molecular dynamics simulation of primary radiation damage in tungsten. *Physical Review B*. 108, 5 (Aug. 2023), 054312. https://doi.org/10.1103/PhysRevB.108.054312.

[28] 2012. The interface is still the device. *Nature Materials*. 11, 2 (Feb. 2012), 91–91. https://doi.org/10.1038/nmat3244.

[29] Kim, S.Y., Sun, Z., Roy, J., Wang, X., Chen, Z., Appenzeller, J. and Wallace, R.M. 2024. Fundamental Understanding of Interface Chemistry and Electrical Contact Properties of Bi and $MoS_2$. *ACS Applied Materials & Interfaces*. 16, 40 (Oct. 2024), 54790–54798. https://doi.org/10.1021/acsami.4c10082.

[30] Oosterbaan, W.D., Bolsée, J., Wang, L., Vrindts, V., Lutsen, L.J., Lemaur, V., Beljonne, D., McNeill, C.R., Thomsen, L., Manca, J.V. and Vanderzande, D.J.M. 2014. On the Relation between Morphology and FET Mobility of Poly(3-alkylthiophene)s at the Polymer/$SiO_2$ and Polymer/Air Interface. *Advanced Functional Materials*. 24, 14 (Apr. 2014), 1994–2004. https://doi.org/10.1002/adfm.201303298.

[31] Xie, J. et al. 2024. Low Resistance Contact to P-Type Monolayer $WSe_2$. *Nano Letters*. 24, 20 (May 2024), 5937–5943. https://doi.org/10.1021/acs.nanolett.3c04195.

[32] Kim, D.H., Kwak, S.J., Jeong, J.H., Yoo, S., Nam, S.K., Kim, Y. and Lee, W.B. 2021. Molecular Dynamics Simulation of Silicon Dioxide Etching by Hydrogen Fluoride Using the Reactive Force Field. *ACS Omega*. 6, 24 (June 2021), 16009–16015. https://doi.org/10.1021/acsomega.1c01824.

[33] Perebeinos, V. and Tersoff, J. 2014. Carbon Nanotube Deformation and Collapse under Metal Contacts. *Nano Letters*. 14, 8 (Aug. 2014), 4376–4380. https://doi.org/10.1021/nl5012646.